 \documentclass[11pt]{article}
 \usepackage[top=1in,bottom=1in,left=1.5in,right=1.5in]{geometry}
  \usepackage{amsmath,amssymb}
  \usepackage{latexsym}% defines $\Box$ for LaTeX2e
 \usepackage[dvips]{pstricks} % PSTricks
  \usepackage{pst-node}
  \usepackage[dvips]{graphicx}

 \newcommand\R{\mathord{\mathbb R}}
 
 \newcommand\C{\mathord{\mathbb C}}

 \newcommand\N{\mathord{\mathbb N}}
 
 \renewcommand\P{\mathord{\mathbb P}}

 \newcommand\W{\mathord{\mathbf W}}
  \renewcommand{\a}{\mathbf{a}}
  \renewcommand{\b}{\mathbf{b}}
  \renewcommand{\i}{\mathbf{i}}
  
  \renewcommand{\c}{\mathbf{c}}

  \newcommand{\gl}{\mathbf{GL}\mathnormal}

  \renewcommand{\u}{\mathbf{u}}
  \renewcommand{\v}{\mathbf{v}}
  
  \newcommand{\w}{\mathbf{w}}
  \newcommand{\x}{\mathbf{x}}
  
  \newcommand{\y}{\mathbf{y}}

  \newcommand{\0}{\mathbf{0}}
  \newcommand{\1}{\mathbf{1}}

 \newcommand\cA{{\cal A}}

 \newcommand\cS{{\cal S}}

 \newcommand\rH{{\rm H}}

 \newcommand\rS{{\rm S}}

  \newcommand{\lan}{\langle}
  \newcommand{\ran}{\rangle}
  \newcommand{\an}[1]{\lan#1\ran}
  \def\diag{\mathop{{\rm diag}}\nolimits}
  \newcommand{\hs}{\hspace*{\parindent}}
  \newcommand{\proof}{\hs \textbf{Proof.\ }}
  
  \newcommand{\tr}{\mathop{\mathrm{tr}}\nolimits}

  \newcommand{\trans}{^\top}
  \newcommand{\qed}{\hspace*{\fill} $\Box$\\}

  \newcommand{\dist}{\mathrm{dist}}

  \renewcommand{\rS}{\mathrm{S}}

  \newcommand{\adj}{\mathrm{adj\;}}

  \newtheorem{theo}{\bfseries \hs Theorem}[section]

  \newtheorem{prob}[theo]{\bfseries \hs Problem}
  \newtheorem{lemma}[theo]{\bfseries \hs Lemma}
  \newtheorem{corol}[theo]{\bfseries \hs Corollary}
  
  \newtheorem{con}[theo]{\bfseries \hs Conjecture}

  \numberwithin{equation}{section} % Automatically number equations within sections

 \renewcommand\det{{\rm det\;}}
 \renewcommand\dim{{\rm dim\;}}

\begin{document}
\title{On Schr\"odinger's bridge problem}
 \author{
  Shmuel Friedland\footnotemark[1]
 }
 \renewcommand{\thefootnote}{\fnsymbol{footnote}}

 \footnotetext[1]{
 Department of Mathematics, Statistics and Computer Science,
 University of Illinois at Chicago, Chicago, Illinois 60607-7045,
 USA, \texttt{friedlan@uic.edu}.   }

 \renewcommand{\thefootnote}{\arabic{footnote}}
 \date{September 3, 2016 }
 \maketitle

 \begin{abstract}
In the first part of this paper we generalize the result of Georgiou-Pavon that a positive square matrix can be scaled uniquely to a column stochastic
matrix which maps a given positive probability vector to another given positive probability vector.  In the second part of this paper
we prove that a positive quantum channel can be scaled to another positive quantum channel which maps a given
positive definite density matrix to another given positive definite density matrix using Brower's fixed point theorem.
This result proves the Georgiou-Pavon  conjecture for two positive definite density matrices in their recent paper \cite{GP15}.  We show uniqueness of fixed points for certain  two positive definite density matrices.

 \end{abstract}


\section{Introduction}
The classical Schr\"odinger bridge problem, studied by Schr\"odinger in \cite{Sch31, Sch32}, seeks the most likely probability law for a diffusion process, in path space, that matches marginals at two end points in time.  
The discrete version of Schr\"odinger's bridge problem for Markov chains, \emph{as I understand}, can be stated succinctly  as follows.   Let $A$ be an $n\times n$ a stochastic matrix.  That is, $A$ is a nonnegative matrix, and $A\trans \1=\1,\1=(1,\ldots,1)\trans$.  (Sometimes $A$ is called \emph{column} stochastic.)
Let $\a,\b$ be two positive (column) probability vectors.  Does there exists a scaling of $A$, i.e., $B=D_1AD_2$, where $D_1,D_2$ are two $n\times n$
diagonal matrices with positive diagonal entries, such that $B$ is stochastic and $B\a=\b$?  See \cite{GP15} for the motivation of this problem and \S\ref{sec:schrodstoc}.

For $\a=\b=\1$ this problem is equivalent to the well known problem: when a nonnegative $A\in\R^{n\times n}, A\ge 0$,
can be scaled to a doubly stochastic matrix?  This problem was answered by Sinkhorn \cite{Sin64}.  Namely, it is always possible to scale a matrix $A$ to a doubly stochastic
if all the entries of $A$ are positive, i.e. $A>0$.  In that case $D_1,D_2$ are unique (up to : $tD_1, t^{-1}D_2$ for $t>0$).  The unique scaling of a fully indecomposable 
$A\ge 0$ to a doubly stochastic matrix was proven in \cite{PM65, BPS66, SK67}.   ($A\ge 0$  fully indecomposable if $PAQ$ is irreducible for any pair of permutation matrices $P,Q$.)
Necessary and sufficient conditions for case $\a=\1$ was given in \cite{Bru68,Men68}.
 Theorem 3.2 in \cite{FK75} proves the existence of unique scaling of a fully indecomposable $A$
for $\a=\b$.  

In a recent paper Georgiou and Pavon \cite{GP15} proved that Schr\"odinger's bridge problem
is uniquely solvable for $A>0$.  Their proof is based on the strict contraction of a corresponding map with respect to the Hilbert metric.
In the first part of this paper we give a different proof of this result.
We also give a generalization of Schr\"odinger's bridge problem to a nonnegative $A$ with no zero column and $AA\trans$ irreducible.
Georgiou and Pavon considered in \cite{GP15} an analog of Schr\"odinger 's bridge problem for quantum channels. The simplified version of this problem is:
Denote by $\C^{n\times n}$ the set of $n\times n$ complex valued matrices, and by $\rH_n\supset \rH_{n,+}\supset \rH_{n,++}$  the subsets of hermitian, positive semi-definite, and positive definite matrices, respectively, in $\C^{n\times n}$.
Let $\rH_{n,+,1}\supset \rH_{n,++,1}$ be the subset  positive semi-definite and positive definite matrices of trace $1$, respectively, in $\rH_{n,+}$. 
Let $\gl(n,\C)\subset \C^{n\times n}$ be the group of invertible matrices.
Recall that $Q:\C^{n\times n}\to \C^{n\times n}$ is called a completely positive operator if
\begin{equation}\label{defcp}
Q(X)=\sum_{i=1}^k A_i X A_i^*, \quad A_i\in\C^{n\times n},\; i=1,\ldots,k, \quad X\in\C^{n\times n}.
\end{equation}
Assume that $Q$ is completely positive.  $Q$ is called positive if $Q(\rH_{n,+,1})\subset \rH_{n,++}$.  $Q$ is called quantum channel if 
\begin{equation} 
\sum_{i=1}^k A_i^* A_i=I_n.  \label{qccond}
\end{equation}
Note that a quantum channel is trace preserving, which is equivalent to $Q(\rH_{n,+,1})\subseteq \rH_{n,+,1}$.  
$R:\C^{n\times n}\to \C^{n\times n}$ is called a scaling of $Q$ if 
\begin{equation}\label{rescdef}
R(X)=SQ(TXT^*)S^*,  \textrm{ for fixed } S,T\in \gl(n,\C) \textrm{ and all } X\in \C^{n\times n}.
\end{equation}
The simplified version Schr\"odinger's bridge problem for quantum channels is:
\begin{prob}\label{SBPQC}  Let $Q:\C^{n\times n}\to \C^{n\times n}$ be a positive quantum channel.  Assume that $\alpha,\beta\in\rH_{n,++,1}$.
Does there exists a scaling of $Q$ to a quantum channel $R$ that satisfies $R(\alpha)=\beta$.
\end{prob}

Problem \ref{SBPQC} for $\alpha=\beta=\frac{1}{n}I_n$ was solved by Gurvits \cite{Gur04} and in \cite{GP15} by different methods.
(This is the analog of Sinkhorn's theorem.)   Conjecture 1 in \cite{GP15} implies the solution of Problem \ref{SBPQC}.  
In this paper we show that Problem \ref{SBPQC} is solvable using Brouwer's fixed point theorem, similar to the methods in \cite{GP15}.   This result yields a solution of the Georgiou-Pavon conjecture for two positive definite density matrices.
We also show that for $\alpha,\beta$ in some 
neighborhood of $\frac{1}{n}I_n$, depending on $Q$, the scaling of $Q$ is "unique" in  certain sense.  That is, the corresponding map has a unique 
fixed point.  

We now summarize briefly the contents of this paper.   In \S\ref{sec:schrodstoc} we discuss Schr\"odinger's bridge problem for stochastic matrices and its generalization to certain nonnegative matrices.  \S\ref{sec:prelim} discusses some known results on quantum channels that are used in the next sections.  In \S\ref{sec:rescqc} give a solution to Problem \ref{SBPQC} using Brower's fixed point theorem.  We also show that a solution to Problem \ref{SBPQC} is equivalent to \cite[Conjecture 1]{GP15} for two positive definite density matrices.  In \S\ref{sec:uniq} we show that the map constructed to solve Problem \ref{SBPQC} has a unique fixed point if the two density matrices are in the neighborhood of the uniform density matrix. 

\section{Schr\"odinger's bridge problem for stochastic matrices}\label{sec:schrodstoc}
Denote $\R_+=[0,\infty), \R_{++}=(0,\infty)$.  For column vectors $\u=(u_1,\ldots,u_n)\trans, \v=(v_1,\ldots,v_n)\trans$ let $\u\circ\v=(u_1v_1,\ldots,u_nv_n)\trans, D(\u)=\diag(u_1,\ldots,u_n)\in\R^{n\times n}$.

\noindent Schr\"odinger's bridge problem is stated as follows \cite{GP15}.  Let $A=[a_{ij}]\in \R^{n\times n}_{++}$, and let
two probability vectors $\a,\b\in\R_{++}^n$ be given.  Do there exist $\u,\v,\x,\y\in \R_{++}^n$ such that
\begin{equation}\label{GPcondit}
\v=A\trans \u,\quad \y=A \x,\quad \a=\v\circ\x,\quad \b=\u\circ\y.
\end{equation}
A straightforward calculation yields that if \eqref{GPcondit} holds then the matrix

\noindent $B=D(\u)A D(\v)^{-1}$ satisfies:
\begin{equation}\label{schrodmatcond}
B\trans \1=\1, \; B\a=\b, \quad B\in\R_{++}^{n\times n},\;\a,\b\in\R^n_{++}, \1\trans \a=\1\trans\b=1.
\end{equation}
Vice versa, if $B=D(\u)AD(\v)^{-1}$ satisfies the above equation then \eqref{GPcondit} holds, where $\x,\y$ are determined uniquely by the last
two conditions of \eqref{GPcondit}.  

Note that any $A\in \R^{n\times n}_{++}$ is uniquely scaled from the right to a stochastic matrix.  That is there exists a unique $D_2=D(A\trans \1)^{-1}$ such
that $AD_2$ is stochastic.  Hence without loss of generality we can assume that $A\in\R^{n\times n}_{++}$ is stochastic.

It is shown in \cite[Theorem 3.2]{FK75}, that for given fully indecomposable $A\in\R^{n\times n}_+$ and $\u,\v\in\R_{++}^n$ there exists a unique scaling of $A$ such that
$D_1AD_2\u=\u$ and $D_2A\trans D_1\v=\v$.   Choose $\a=\b=\u$ and $\v=\1$ to deduce the solution of Schr\"odinger's bridge
problem in this case.  It is straightforward to show that a solution of Schr\"odinger's bridge problem for $\a=\b$ implies the above cited result in \cite{FK75} by considering the matrix
$D(\v)^{-1}D_1AD_2D(\v)$ and $\a=\b=\u\circ\v$.

Denote $\P\R_{++}^n$ the projective space associated with the open cone of positive vectors in $\R^n$.  That is,  $\P\R_{++}^n$ is the set of open rays $R(\u)=\{t\u, \; t>0,
\u\in\R_{++}^n\}$.  Denote by $\Pi_n\subset\R^n$ the simplex of probability vectors.
Clearly, $\P\R^n_{++}$ is isomorphic to the interior of $\Pi_n$, denoted as $\Pi_{n,++}= \Pi_n\cap \R^n_{++}$ .   A one point compactification of $\Pi_{n,++}$ is the identification of the points $\partial \Pi_{n,++}$, the boundary of $\Pi_{n,++}\subset \R^n$, to one point, denoted as $\infty$.  Denote by $\widehat{\Pi_n}=\Pi_{n,++}\cup\{\infty\}$ the above one point compactification of $\Pi_n$.  It is well known that $\widehat {\Pi_n}$ is homeomorphic to the $n-1$ dimensional sphere $\rS^{n-1}=\{\x\in\R^n,\;\x\trans \x=1\}$.  Moreover, $\rS^{n-1}$ can be viewed as one point compactification of $\R^{n-1}$: $\widehat{\R^{n-1}}=\R^{n-1}\cup\{\infty\}$.  Similarly, one point compactification of $\P\R_{++}^n$, $\widehat{\P\R_{++}^n}=\P\R_{++}^n\cup\{\infty\}$, is homeomorphic to $\rS^{n-1}$, equivalently homeomorphic to $\widehat{\Pi_n}$.

Denote by $\Sigma_n\subset \R_+^{n\times n}$ the convex set of stochastic matrices.
Let $\Sigma_{n,++}=\Sigma_n\cap \R_{++}^{n\times n}$ be the interior of $\Sigma_n$.

Assume that $A\in\R^{n\times n}_{+}$ is a matrix with no zero column.  Let 
\begin{equation}\label{defPhimapmatr}
\Phi_A:\R_{++}^n\to \Sigma_n, \quad \Phi_A(\x)=D(\x)AD(A\trans \x)^{-1} \textrm{ for }\x\in\R_{++}^n.
\end{equation}
Clearly, for each $t>0$ and $\x\in\R^n_{++}$ we have that $\Phi_A(t\x)=\Phi_A(\x)$.
Hence can be viewed as smooth map $\tilde \Phi_A:\P\R_{++}^n\to \Sigma_n$.  Clearly,
$\tilde \Phi_A$ can be viewed also as the restriction of $\Phi_A$ to $\Pi_{n,++}$.
Note that if $A>0$ then $\Phi_A(\R_{++}^n)\subset \Sigma_{n,++}$.
 
The following theorem generalizes \cite[Theorem 3]{GP15}.
\begin{theo}\label{GPthm}  Let $A\in\R^{n\times n}_+$ and $\a\in\Pi_{n,++}$ be given.  Assume that $A$ does not have a zero column or zero row.
Consider the map
\begin{equation}\label{defPhia}
\Phi_{A,\a}:\Pi_{n,++}\to \Pi_{n,++}, \quad \Phi_{A,\a}(\x)=\Phi_A(\x)\a \textrm{ for }\x\in  \Pi_{n,++}.
\end{equation}
\begin{enumerate}
\item Assume that $A$ is a positive matrix.  Then the map $\Phi_{A,\a}$ extends to a continuous map of $\Pi_n$.  It maps the boundary of $\Pi_n$ to its boundary.  Furthermore, 
$\Phi_{A,\a}$ is a self-diffeomorphism of $\Pi_{n,++}$.
\item Assume that $AA\trans$ is an irreducible matrix.  Then $\Phi_{A,\a}$ is a diffeomorphism of $\Pi_{n,++}$ and $\Phi_{A,\a}(\Pi_{n,++})$.
\end{enumerate}
\end{theo}
\proof  As $A$ is a nonnegative matrix with no zero column the map \eqref{defPhia} is a well defined smooth map.  As $A$ does not have a zero row, $\Phi_{A}(\x)$ does not have 
zero row for $\x\in\Pi_{n,++}$.  Therefore $\Phi_{A}(\x)\a\in\Pi_{n,++}$.  Hence $\Phi_{A,\a}$ is a self-smooth map of $\Pi_{n,++}$, i.e. $\Phi_{A,\a}(\Phi_{n,++})\subseteq \Pi_{n,++}$.

\noindent
\emph{1}.  Assume that $A>0$.  Then  $\Phi_A:\Pi_n\to \Sigma_n$ is continuous.
In particular, $\Phi_{A}(\x)\a\in \Pi_n$ for each $\x\in\Pi_n$.  Therefore $\Phi_{A,\a}$ is a continuous map of
$\Pi_n$ to itself.  Let $\x=(x_1,\ldots,x_n)\trans\in \partial \Pi_n$.  Thus $x_i=0$ for some $i$.  Hence the $i-th$ row of $\Phi_{A}(\x)$ is zero.  Therefore the $i$-th coordinate of $\Phi_A(\x)\a$ is zero.  Thus $\Phi_{A}(\x)\a\in\partial \Pi_{n,++}$.  Hence $\Phi_{A,\a}(\partial \Pi_{n,++})\subseteq \partial \Pi_{n,++}$.  These results yield that $\Phi_{A,a}:\Pi_n\to \Pi_n$
induces a continuous map $\widehat{\Phi_{A,\a}}:\widehat{\Pi_n}\to \widehat{\Pi_n}$. 
 More precisely the map $\Phi_{A,\a}:\Pi_{n,++}\to \Pi_{n,++}$ is a proper map.
That is, given a sequence $\x_m\in \Pi_{n,++}, m\in\N$ which converges to $\partial \Pi_{n,++}$ then the sequence $\Phi_{A,\a}(\x_m), m\in \N$ converges to  $\partial \Pi_{n,++}$.

We now show that $\Phi_{A,\a}: \Pi_{n,++}\to \Pi_{n,++}$ is a local diffeomorphism.  Assume that $\x\in\Pi_{n,++}$.
Then the neighborhood of $\x$ in $\Pi_n$ are points of the form 
$\x+t\w, \w\trans \w=1, \1\trans \w=0, |t|<\varepsilon$ for some small $\varepsilon>0$.  Equivalently, let $\W\subset \R^n$ be the subspace of all vectors orthogonal to $\1$.  Then the neighborhood of $\x$ is diffeomorphic to an open ball of radius $\varepsilon>0$ centered at $\0$ in $\W$.

Suppose first that $A$ is stochastic.  
Assume that $\x=\frac{1}{n}\1$. 
Thus
\begin{eqnarray*}
&&A\trans(\x+t\w)=\x+t A\trans \w, (A\trans(\x+t\w))_i^{-1}=n(\1-n t A\trans\w)+O(t^2),\\ 
&&D(A\trans(\x+t\w))^{-1}=n(I -nt D(A\trans\w))+O(t^2), \\
&&\Phi_A(\x+t\w)=A+nt(D(\w)A-AD(A\trans\w))+O(t^2).
\end{eqnarray*}
Assume that $A\a=\b$.  Then
\[\Phi_{A,\a}(\x+t\w)=\b +nt(D(\w)\b-AD(A\trans\w)\a)+O(t^2)=\b + nt(D(\b)-A D(\a)A\trans)\w +O(t^2).\]
Let 
\begin{equation}\label{defFAa}
F(A,\a)=D(\b)-AD(\a)A\trans=D(A\a)-AD(\a)A\trans.
\end{equation}
Clearly $F(A,\a)$ is symmetric matrix satisfying 
\[F(A,\a)\1=D(\b)\1-AD(\a)A\trans\1=\b-AD(\a)\1=\b-A\a=\0.\]
Hence $\W$ is an invariant subspace of $F(A,\a)$.  Furthermore, $nF(A,\a)|\W$ is the Jacobian of $\Phi_{A,\a}$ at $\x=\frac{1}{n}$.

We claim that the $n-1$ eigenvalues of $F(A,\a)|\W$ are positive.  This claim follows from the observation that $F(A,\a)$ a symmetric irreducible singular M-matrix \cite[\S6.6]{Frbk}.
Indeed, the matrix $rI-F(A,\a)>0$ for $r\gg 1$.   Clearly $(rI-F(A,\a))\1=r\1$.  Hence $r$ is the spectral radius of $(rI-F(A,\a))$,  $r$ is a simple eigenvalue
of  $rI-F(A,\a)$ and all other eigenvalues of $rI-F(A,\a)$ are strictly less than $r$.  Hence $-F(A,\a)$ is a singular symmetric matrix, which has $n-1$ negative eigenvalues.  
In particular $F(A,\a)|\W$ is an invertible transformation.  Hence $\Phi_{A,\a}$ is a local diffeomorphism at $\x=\frac{1}{n}\1$.  
For a general $\x\in\Pi_{n,++}$ if follows that the Jacobian of $\Phi_{A,\a}$ at $\x$ is $nF(\Phi_{A}(\x),\a)|\W$.  Hence $\Phi_{A,a}$ is a local diffeomorphism on $\Pi_{n,++}$.

 As $\Phi_{A,\a}: \Pi_{n,++}\to  \Pi_{n,++}$ is a proper map and a local diffeomorphism it follows that
$\Phi_{A,\a}$ is proper cover of $\Pi_{n,++}$ by $\Pi_{n,++}$.  As $\Pi_{n,++}$ is simply connected we deduce that $\Phi_{A,\a}$ is a diffeomorphism of  $\Pi_{n,++}$.
(One can also prove this fact by using the degree theory \cite{Mil}.)

\noindent
\emph{2}.  It is left to discuss the theorem where $A$ is a nonnegative matrix with some zero entries, with no zero columns and $A A\trans$ is irreducible.
(This condition yields that $A$ has no zero rows.)  We claim first that $\Phi_{A,\a}$ is a local diffeomorphism.  As above, the Jacobian of  $\Phi_{A,\a}$ at $\x\in\Pi_{n,++}$
is $n$ times the restriction of
\[F(\Phi_{A}(\x),\a)=D(\Phi_{A}(\x)\a)-\Phi_{A}(\x)D(\a)\Phi_A(\x)\trans\]
to $\W$.  Clearly, $F(\Phi_{A}(\x),\a)\1=\0$.  We claim that  $F(\Phi_{A}(\x),\a)$ is an irreducible singular M-matrix.  Indeed, $(i,j)$ off-diagonal entry of 
$-F(\Phi_{A}(\x),\a)$ is positive if and only if the $(i,j)$ entry of $A A\trans$ is positive.  Since $A A\trans$ is an irreducible matrix it follows that $F(\Phi_{A}(\x),\a)$ a symmetric irreducible singular $M$-matrix.   Hence the eigenvalues of $F(\Phi_{A}(\x),\a)|\W$ are positive, and  the Jacobian of  $\Phi_{A,\a}$ at $\x$ is invertible.  

It is left to show that $\Phi_{A,\a}:\Pi_{n,++}\to \Phi_{A,\a}(\Pi_{n,++})$ is one-to-one.  Assume to the contrary that there exists $\x_1,\x_2\in \Pi_{n,++}, \x_1\ne \x_2$
such that $\b=\Phi_{A,\a}(\x_1)=\Phi_{A,\a}(\x_2)$.  Since $\Phi_{A,\a}$ is a local diffeomorphism the following conditions hold.
There exists a closed ball  of a small radius $\varepsilon$  centered at $\0$: $B(\0,\varepsilon)\subset \W$, such that
\[\x_1+B(\0,\varepsilon),\x_2+B(\0, \varepsilon)\subset \P_{n,++}, \quad  (\x_1+B(\0,\varepsilon))\cap (\x_2+B(\0,\varepsilon))=\emptyset,\] and there exists  
$\varepsilon'>0$ such that 
\[\Phi_{A,\a}(\x_1+B(\0,\varepsilon))\cap \Phi_{A,\a}(\x_2+B(\0,\varepsilon)) \supset \b+B(\0,\varepsilon').\]
Assume that $t>0$ is small and let $A(t)=A+t\1\1\trans>0$.  So $A(t)$ is a positive perturbation of $A$.  Therefore $\Phi_{A(t),\a}$ is a perturbation of the 
map $\Phi_{A,\a}$ on the closed sets $\x_1+B(\0,\varepsilon), \x_2+B(\0,\varepsilon)$.  The above condition yields that
\[\Phi_{A(t),\a}(\x_1+B(\0,\varepsilon))\cap \Phi_{A(t),\a}(\x_2+B(\0,\varepsilon)) \supset \b+B(\0,\varepsilon_1)\]
for some $\varepsilon_1\in (0,\varepsilon')$ for some very small and positive $t$.
This contradicts that $\Phi_{A(t),\a}$ is a self-diffeomorphism of $\Pi_{n,++}$.\qed

We now discuss briefly the results of Theorem \ref{GPthm}.  Clearly, part \emph{1} of this theorem is equivalent to the result of Georgiou-Pavon for positive matrices \cite{GP15}.
We now consider  part \emph{2} of this theorem.  First note part \emph{2} fails if $A A\trans$ is not irreducible.  Indeed, suppose that $A$ is a permutation matrix.
Then $\Phi_{A}(\x)=A$ for each $\x\in\Pi_{n,++}$.  Therefore the map $\Phi_{A,\a}$ is a constant map.  Note that $A A\trans=I$.

Suppose that $A$ is fully indecomposable.  This is equivalent to the statement that $A=PBQ$, where $P$ and $Q$ are permutation matrices, $B=D+C$, such that $D$ is a diagonal matrix with positive diagonal entries and $C\ge 0$ is an irreducible matrix \cite{BPS66}.  Then $AA\trans=P(D^2+DC\trans+CD +CC\trans)P\trans$ is irreducible.  It is easy to given an example of a fully indecomposable $A$ such that $\partial\Phi_{A,\a}(\Pi_{n,++})\cap \Pi_{n,++}\ne \emptyset$.  

Assume that 
\begin{equation}\label{exdfimat}
A=\left[\begin{array}{ccc}1&1&0\\0&1&1\\1&0&1\end{array}\right], \;\x(t)=\left[\begin{array}{c}tc_1\\tc_2\\1-tc_1-tc_2\end{array}\right],\; c_1, c_2,t>0,\; c_1+c_2=1
\end{equation}
 and $t\searrow 0$.  Then
\begin{equation}\label{defBlim}
\lim_{t\searrow 0} \Phi_{A}(\x(t))=B=\left[\begin{array}{ccc}0&c_1&0\\0&c_2&0\\1&0&1\end{array}\right], \quad \lim_{t\searrow 0} \Phi_{A,\a}(\x(t))=B\a.
\end{equation}
Hence $\0<B\a\in \partial\Phi_{A,\a}(\Pi_{3,++})$.

 Theorem 3.2 in \cite{FK75} is equivalent to the statement that for a fully indecomposable $A$ and $\a\in \Pi_{n,++}$ one has $\a\in \Phi_{A,\a}(\Pi_{n,++})$.  

The results in 
\cite{Bru68,Men68} can be stated as follows.  Given a fully indecomposable matrix $A=[a_{ij}]\in\R_+^{n\times n}$ and $\b,\c\in\Pi_{n,++}$ there exist two diagonal matrices $D_1,D_2\in\R_+^{n\times n}$, with positive diagonal entries, such that $D_1AD_2\1=n\b, D_2A\trans D_1 \1=n\c$ if and only if the following condition holds.
There exists a matrix $C=[c_{ij}]\in\R^{n\times n}_+$ having the same pattern, i.e. $a_{ij}>0\iff c_{ij}>0$ for $i,j=1,\ldots,n$, such that $C\1=n\b, C\trans\1=n\c$.
Hence for a fully indecomposable matrix $A$ one has $\b\in  \Phi_{A,\frac{1}{n}\1}(\Pi_{n,++})$ if and only if there exists a stochastic $C\in\R_+^{n\times n}$ having the same pattern as $A$ such that $C\1=n\b$.

Let $A$ and $B$ be defined as in \eqref{exdfimat} and \eqref{defBlim} respectively.  Clearly, $A$ is fully indecomposable.  Let $\b=B(\frac{1}{3}\1)=(\frac{c_1}{3},\frac{c_2}{3}, \frac{2}{3})\trans$.  It is straightforward to show that there is no stochastic matrix $C\in\R_+^{3\times 3}$ with the same pattern as $A$ such that $C\1=3\b$.  So $\b\notin \Phi_{A,\frac{1}{3}\1}(\Pi_{3,++})$ as we claimed.

\section{Preliminary results on quantum channels}\label{sec:prelim}
For $X\in\rH_n$ arrange the eigenvalues of $X$ in a non increasing  order $\lambda_1(X)\ge\cdots\ge\lambda_n(X)$.
Let trace $X$ be the trace of X: $\tr X=\sum_{i=1}^n \lambda_i(X)$.  Denote by $\|X\|_F=\sqrt{\tr F^2}=\sqrt{\sum_{i=1}^n \lambda_i(X)^2}$ the Frobenius norm of $X$.
For two real number $a\le b$ denote by 
\[\rH_n(a,b)=\{X\in \rH_n, \;a\le \lambda_n(X), \lambda_1(X)\le b\}.\]
It is well known that $\rH_n(a,b)$ is a convex set.  (Recall that $\lambda_1(X)$ and $-\lambda_n(X)$ are convex functions on $\rH_n$ \cite{Frbk}.)  
Let
\begin{equation}\label{defnHn1ab}
\rH_n(a,b,1):=\{X\in \rH_n(a,b), \;\tr X=1\}.
\end{equation}
Note that $\rH_n(a,b,1)\ne \emptyset $ if and only if $a\le \frac{1}{n}\le b$.  Furthermore, given $a\le \frac{1}{n}$ then for each $X\in\rH_n(a,b,1)$ it follows that
$\lambda_1(X)\le 1-(n-1)a$.  
Clearly, for 
\begin{equation}\label{abcond}
0<a\le \frac{1}{n}\le b\le 1-(n-1)a
\end{equation} 
the convex set $\rH_n(a,b,1)$ is a convex compact set of $\rH_{n,++,1}$.
For a completely positive $Q$  let 
\begin{equation}\label{defabQ}
a(Q)=\min_{X\in\rH_{n,+,1}} \lambda_n(Q(X)), \quad b(Q)=\max_{X\in\rH_{n,+,1}}\lambda_1(Q(X)).
\end{equation}
Clearly, $Q(\rH_{n,+,1})\subseteq \rH_n(a(Q),b(Q))$.
Thus $Q$ is positive if and only if $a(Q)>0$. For a positive quantum channel $Q$ the constants $a(Q)$ and $b(Q)$ satisfy \eqref{abcond} and 
$Q(\rH_{n,+,1})\subseteq \rH_n(a(Q),b(Q),1)$. 

Let $\|X\|=\max_{\|\x\|=1} \|X\x\|$ be the spectral norm of $X\in \C^{n\times n}$.  Recall that 
\begin{equation}\label{Xnormid}
\|X\|=\lambda_1(X), \;\lambda_n(X)=\frac{1}{\lambda_1(X^{-1})}=\frac{1}{\|X^{-1}\|},\textrm{ for } X\in \rH_{n,++}.
\end{equation}
Furthermore, for each $X\in\rH_{n,+}$ there exists a unique positive semi-definite matrix $X^{\frac{1}{2}}\in \rH_{n,+}$, the square root of $X$, so that $(X^{\frac{1}{2}})^2=X$.

\begin{lemma}\label{Dalphbd}
Assume that $\alpha\in \rH_{n,+}\setminus\{0\}$.   Let  $D_{\alpha}:\rH_{n,++} \to\rH_{n,+,1}$ be given by
\begin{equation}\label{defDalph}
D_{\alpha}(X)=\frac{1}{\tr X^{-1}\alpha}X^{\frac{-1}{2}}\alpha X^{-\frac{1}{2}}.
\end{equation}
If $\alpha\in \rH_{n,++}$ then $D_{\alpha}(\rH_{n,++})\subseteq \rH_{n,++,1}$.
Assume that $a,b$ satisfy \eqref{abcond}.  Then
\begin{eqnarray}\label{Dalphbd1}
&&D_{\alpha}(\rH(a,b,1))\subseteq \rH(c,d,1), \\
&&c=\frac{a\lambda_n(\alpha)}{a\lambda_n(\alpha)+(n-1)b\lambda_1(\alpha)},\; d=\frac{b\lambda_1(\alpha)}{b\lambda_1(\alpha)+(n-1)a\lambda_n(\alpha)}.\notag
\end{eqnarray}
\end{lemma}
\proof  Clearly, for $X\in\rH_{n,++}$ and $\alpha\in\rH_{n,+}\setminus\{0\}$ we have that $X^{\frac{-1}{2}}\alpha X^{-\frac{1}{2}}\in \rH_{n,+}\setminus\{0\}$.  
Hence $D_{\alpha}(\rH_{n,++})\subseteq \rH_{n,+,1}$.  Furthermore if $\alpha\in \rH_{n,++,1}$ then $D_{\alpha}(\rH_{n,++,1})\subseteq \rH_{n,++,1}$.

Assume that $\alpha\in\rH_{n,++,1}$.  Clearly,
\[\lambda_1(X^{\frac{-1}{2}}\alpha X^{-\frac{1}{2}})=\|X^{\frac{-1}{2}}\alpha X^{-\frac{1}{2}}\|\le \| X^{-\frac{1}{2}}\|^2\|\alpha\|=
\frac{\lambda_1(\alpha)}{\lambda_n(X)}\le \frac{\lambda_1(\alpha)}{a}.\]
As $\lambda_n(X^{\frac{-1}{2}}\alpha X^{-\frac{1}{2}})=\frac{1}{\lambda_1(X^{\frac{1}{2}}\alpha^{-1} X^{\frac{1}{2}})}$ it follows that 
\[\lambda_n(X^{\frac{-1}{2}}\alpha X^{-\frac{1}{2}})\ge \frac{\lambda_n(\alpha)}{\lambda_1(X)}\ge \frac{\lambda_n(\alpha)}{b}.\]
Observe next that if $0<f\le x_n\le\cdots \le x_1\le g$ then
\[\frac{f}{f+(n-1)g}\le \frac{x_n}{\sum_{i=1}^n x_i}\le \ldots\le \frac{x_1}{\sum_{i=1}^n x_i}\le\frac{g}{g+(n-1)f}.\]
Combine the above results to deduce \eqref{Dalphbd1}.  The continuity argument yields
\eqref{Dalphbd1} for $\alpha\in\rH_{n,+,1}$. \qed

For a positive integer $k$ let  $[k]:=\{1,\ldots,k\}$.
Recall that on $\C^{n\times n}$ one has the inner product $\an{X,Y}=\tr XY^*$.
For a completely positive operator $Q$ given by \eqref{defcp} let: 
\begin{equation}\label{defQprime}
Q'(X)=\sum_{i=1}^k A_i^* X A_i, \quad A_i\in\C^{n\times n},\; i\in [k].
\end{equation}
Then $Q'$ is the dual of $Q$.
Indeed, $\an{Q(X),Y}=\an{X,Q'(Y)}$.   Clearly, $Q'$ is completely
positive.  Thus $Q$ is a quantum channel if and only if $Q'(I_n)=I_n$.  $Q$ is called a unital channel if $Q'(I_n)=Q(I_n)=I_n$.
A unital channel is an analog of a doubly stochastic matrices, and sometimes is called doubly stochastic quantum channel \cite{MW09}.

For $A,B\in\rH_n$ we denote $A\succeq B, A\succneqq B, A \succ B$ if $A-B$ is positive semi-definite,
nonzero positive semi-definite, positive definite, respectively.   Clearly, any completely operator $Q:\C^{n\times n}\to \C^{n\times n}$
is order preserving on $\rH_n$, i.e. $X\succeq Y \Rightarrow Q(X)\succeq Q(Y)$.  
Hence $Q'(X)\succeq 0$ if $X\succeq 0$. 

Assume that $X,Y\in \rH_{n,+}$ then $X^{\frac{1}{2}}YX^{\frac{1}{2}}\in\rH_{n,+}$.  Therefore $\tr XY= \tr X^{\frac{1}{2}}YX^{\frac{1}{2}}\ge 0$.
Hence if $U\preceq V$ it follows that $\tr UY\le \tr VY$.
\begin{lemma}\label{posQprime}  
Assume that $Q:\C^{n\times n}$ is a completely positive operator.   Then
\begin{equation}\label{posQprime1}
a(Q)=a(Q'), \quad b(Q)=b(Q').
\end{equation}
In particular $Q$ is positive if and only if $Q'$ is positive. 
\end{lemma}
\proof
From the definition of $a(Q),b(Q)$ given by \eqref{defabQ} it follows that $a(Q)I_n \le Q(X)\le b(Q)I_n$ for each $X\in\rH_{n,+.1}$.
From the definition of $a(Q'),b(Q')$ it follows that there exists $U,V\in\rH_{n,+,1}$ and $\u,\v\in\C^n, \|\u\|=\|\v\|=1$ such that
\[a(Q')=\lambda_n(Q'(U))=\u^*Q'(U)\u, \quad b(Q') =\lambda_1(Q'(V))=\v^* Q'(V)\v.\]
Hence
\begin{eqnarray*}
&&a(Q')=\tr \u\u^* Q'(U)=\tr Q(\u\u^*) U\ge \tr a(Q)I_n U=a(Q),\\  
&&b(Q')=\tr \v\v* Q'(V)=\tr Q(\v\v^*)V\le \tr b(Q)I_n V=b(Q).
\end{eqnarray*}
As $(Q')'=Q$ it follows that $ a(Q)=a((Q')')\ge a(Q')$ and $b(Q')\ge b((Q')')=b(Q)$.  Hence \eqref{posQprime1} holds.
In particular, $a(Q)>0$ if and only if $a(Q')>0$.\qed

Lemma \ref{posQprime} and the arguments of the proof of Lemma \ref{Dalphbd} yield.
\begin{corol}\label{Qprimebounds}  Assume that $Q$ is a positive completely positive operator given by \eqref{defcp}.
Let $Q'$ be given by \eqref{defQprime}.  Denote by $\tilde Q:\rH_{n,+,1}\to \rH_{n,++,1}$ the following nonlinear operator
\begin{equation}\label{defQtilde}
\tilde Q(X)=\frac{1}{\tr Q'(X)} Q'(X), \quad X\in\rH_{n,+,1}.
\end{equation}
Then
\begin{eqnarray}\label{bdsabQtild}
&&\tilde Q(\rH_{n,+,1})\subseteq \rH_n(e,f,1),\\
&&\frac{a(Q)}{a(Q)+(n-1)b(Q)}\le e\le f\le \frac{b(Q)}{b(Q)+(n-1)a(Q)}.  \notag
\end{eqnarray}
\end{corol}
Assume that $Q$ is a unital quantum channel.  In this case $\tilde Q=Q'$.   Lemma \ref{posQprime} yields that in \eqref{bdsabQtild} we can assume that $e=a(Q)$ 
and $f=b(Q)$.

\section{Existence of scaling for positive quantum channel}\label{sec:rescqc}
\begin{theo}\label{existfixpts}  Let $Q:\C^{n\times n}\to \C^{n\times n}$ be a positive quantum channel.  Assume that $\alpha,\beta\in\rH_{n,+,1}$.
Consider the following continuous nonlinear transformation $\Phi_{\alpha,\beta}:\rH_{n,+,1}\to\rH_{n,+,1}$:
\begin{equation}\label{defPhi}
\Phi_{\alpha,\beta}=D_{\alpha}\circ\tilde Q\circ D_{\beta}\circ Q.
\end{equation}
Then $\Phi_{\alpha,\beta}$ has a fixed point $U(\alpha,\beta)\in\rH_{n,+,1}$.  If $\alpha\in\rH_{n,++,1}$ then $U(\alpha,\beta)\in \rH_{n,++,1}$.
\end{theo}
\proof  Let $\Psi_1=D_{\beta}\circ Q, \Psi_2=\tilde Q\circ D_{\beta}\circ Q$.  As $Q$ is positive $Q(\rH_{n,+,1})\subseteq \rH_n(a(Q),b(Q),1)$.
Hence $\Psi_1$ is continuous on $Q(\rH_{n,+,1})$.  Similarly, $\Psi_2(\rH_{n,+,1})\subset \tilde Q(\rH_{n,+,1})\subseteq \rH_n(a(Q),b(Q),1)$. 
Thus $\Phi_{\alpha,\beta}$ is continuous on $\rH_{n,+,1}$.  Brouwer's fixed point theorem yields that $\Phi_{\alpha,\beta}$ has a fixed point $U(\alpha,\beta)\in \rH_{n,+,1}$.

Assume finally that $\alpha\in \rH_{n,++,1}$.  Then $D_{\alpha}(\rH_n(a(Q),b(Q),1))\subset \rH_{n,++,1}$.  Hence $U(\alpha,\beta)\in \rH_{n,++,1}$.
\qed

\begin{theo}\label{schridbridgethm}    Let $Q:\C^{n\times n}\to \C^{n\times n}$ be a positive quantum channel.  Assume that $\alpha,\beta\in\rH_{n,++,1}$.
Then each fixed point of $\Phi_{\alpha,\beta}$, given by \eqref{defPhi}, induces a scaling of the quantum channel $Q$ to a quantum channel $R$ satisfying $R(\alpha)=\beta$.
Vice versa, each scaling of the quantum channel $Q$ to a quantum channel $R$ satisfying $R(\alpha)=\beta$ induces a fixed point of 
$\Phi_{\alpha_1,\beta_1}$, for $\alpha_1=O\alpha O^*, \beta_1=P\beta P^*$ for some unitary $O$ and $P$.
\end{theo}
\proof  Suppose that $\Phi_{\alpha,\beta}(U)=U$ for some $U\in\rH_{n,++.1}$.
Denote $V=Q(U), W=D_{\beta}(V), Z=\tilde Q(W)$.  Then $D_{\alpha}(Z)=U$.  Observe that $Q'(W)=t^{-2} Z$ for some $t>0$.
Let $T= tZ^{-\frac{1}{2}}, S=W^{\frac{1}{2}}$.  Let $R_1:\C^{n\times n}\to \C^{n\times n}$ be given by $R_1(X)=SQ(TXT^*)S^*$.
Clearly, $R_1$ is a scaling of $R$.  Furthermore
\[R_1'(I_n)=T^*Q'(S^*S)T=(tZ^{-\frac{1}{2}})Q'(W)(tZ^{-\frac{1}{2}})=(tZ^{-\frac{1}{2}})(t^{-2} Z)(tZ^{-\frac{1}{2}})=I_n.\]
Hence $R_1$ is a quantum channel. 

We now claim that there exists a unitary $O$ such that
$R_1(\alpha)=O\beta O^*$.  Indeed, 
\[Q(U)=Q(D_{\alpha}(Z))=Q(\frac{1}{\tr Z^{-1}\alpha} Z^{-\frac{1}{2}}\alpha Z^{-\frac{1}{2}})=\frac{1}{t^2{\tr Z^{-1}\alpha}} Q(T \alpha T^*)=V.  \]
Hence 
$R_1(\alpha)=(st)^2 W^{\frac{1}{2}}VW^{\frac{1}{2}}$ for some $s=\sqrt{\tr Z^{-1}\alpha}$.
The equality $W=D_{\beta}(V)$  is equivalent to $W=r^2V^{-\frac{1}{2}}\beta  V^{-\frac{1}{2}}$ for $r=\frac{1}{\sqrt{\tr V^{-1}\beta}}$. That is
\[(W^{\frac{1}{2}})^2=(rV^{-\frac{1}{2}}\beta^{\frac{1}{2}}) (rV^{-\frac{1}{2}}\beta^{\frac{1}{2}})^*
\Rightarrow (rW^{-\frac{1}{2}}V^{-\frac{1}{2}}\beta^{\frac{1}{2}}) (rW^{-\frac{1}{2}}V^{-\frac{1}{2}}\beta^{\frac{1}{2}})^*=I_n\]
Therefore $O:=rW^{-\frac{1}{2}}V^{-\frac{1}{2}}\beta^{\frac{1}{2}}$ is a unitary matrix.   Clearly, $V^{\frac{1}{2}}W^{\frac{1}{2}}=r\beta^{\frac{1}{2}} O^*$.
Finally, 
\[R_1(\alpha)=(st)^2 W^{\frac{1}{2}}VW^{\frac{1}{2}}=(st)^2 (W^{\frac{1}{2}}V^{\frac{1}{2}})(W^{\frac{1}{2}}V^{\frac{1}{2}})^*=(rst)^2 O\beta O^*.\]
As $\alpha,\beta\in\rH_{n,+,1}$ and $R_1$ is a trace preserving, it follows that $(rst)^2=1$.  Hence $R_1(\alpha)=O\beta O^*$.  Define
$R(X)=O^*R_1(X) O$.  Clearly, $R$ a scaling of $Q$, $R$ is a quantum channel, and $R(\alpha)=\beta$.  

Assume now that $R$ is a scaling of the quantum channel $Q$ such that $R$ is a quantum channel, and $R(\alpha)=\beta$.  Let $R$ be given by \eqref{rescdef}.  
Define $W:=s S^*S$, where $s=\frac{1}{\tr S^*S}$.  Let $Z=\tilde Q(W)$.  As $R$ is a quantum channel we deduce that $Z=t (T^*)^{-1}T^{-1}$, where 
$t=\frac{1}{\tr  (T^*)^{-1}T^{-1}}$.   From the above arguments it follows that 
\[Z^{\frac{1}{2}}=t^{\frac{1}{2}}(T^*)^{-1}O^*=t^{\frac{1}{2}}OT^{-1}\] 
for some unitary matrix $O$.  Let $\alpha_1=O\alpha O^*$.  Then
\[U=D_{\alpha_1}(Z)=r^{-2}TO^*\alpha_1 OT^*=r^{-2}T\alpha T^*, \quad  r^{2}=\tr T\alpha T^*. \] 
Then $U=r^{-2}T\alpha T^*$.  As $R(\alpha)=\beta$ it follows that $ V=Q(U)=r^{-2} S^{-1}\beta (S^*)^{-1}$.  Hence 
$V^{\frac{1}{2}}=r^{-1}S^{-1}\beta^{\frac{1}{2}}P^*=r^{-1}P\beta^{\frac{1}{2}}(S^*)^{-1}$ for some unitary
$P$.   Let $\beta_1=P\beta P^*$.  Then 
\[D_{\beta_1}(V)=q S^*\beta^{-\frac{1}{2}}P^*\beta_1P\beta^{-\frac{1}{2}}S=qS^*S, \quad q=\frac{1}{\tr S^* S}.\] 
Therefore $D_{\beta_1}(V)=W$, and $U$ is a fixed point of $\Phi_{\alpha_1,\beta_1}$.\qed

Note that if $Q$ is a quantum channel that satisfies $Q(\alpha)=\beta$, then the scaled channel $R$ given by \eqref{rescdef}, with $S$ and $T$ unitary,
is also a quantum channel with $R(T^*\alpha T)=S\beta S^*$.  This observation explains the second part of Theorem \ref{schridbridgethm}.

Theorem \ref{schridbridgethm} proves the Georgiou-Pavon conjecture \cite[Conjecture 1]{GP15} for two positive definite density matrices:
\begin{con}\label{GPcon}(Georgiou-Pavon) Given a positive quantum channel $Q:\C^{n\times n}\to
\C^{n\times n}$ and two density matrices $\rho_0, \rho_T$ then there exists $\phi_0,\phi_T,\hat \phi_0,\hat \phi_T\in \rH_{n,++}$ such that
\begin{eqnarray}\label{GPcon1}
&&Q(\phi_T)=\phi_0, \quad Q'(\hat \phi_0)=\hat\phi_T,\\
&&\rho_0=\chi_0\hat \phi_0\chi_0^*, \quad \rho_T=\chi_T\hat \phi_T \chi_T^*, \label{GPcon2}\\
&&\phi_0=\chi_0^*\chi_0, \quad \phi_T=\chi_T^*\chi_T.\label{GPcon3}
\end{eqnarray}
Furthermore, $\chi_0$ and $\chi_T$ can be chosen to be in $\rH_{n,+}$.
\end{con}

It is shown in \cite{GP15} that this conjecture holds for $\rho_0=\rho_T=\frac{1}{n}I_n$, see also \cite{Gur04}, and for rank one matrices $\rho_0,\rho_T\in\rH_{n,+,1}$.\\

\noindent
\textbf{Proof of Georgiou-Pavon conjecture for $\rho_0,\rho_T\in\rH_{n,++.1}$ and nonhermitian $\chi_0,\chi_T$}.  Let $\alpha=\rho_T$ and $\beta=\rho_0$.  Theorem \ref{existfixpts} implies that
$\Phi_{\alpha,\beta}$ has a fixed point in $\rH_{n,++,1}$.  In the proof of Theorem \ref{schridbridgethm} we constructed the quantum channel $R_1(X)=SQ(TXT^*)S^*$
such that $R_1(\alpha)=O\beta O^*$ for a corresponding unitary $O$.  Furthermore $S,T\in \rH_{n,++}$.   Thus
\[Q(T\alpha T)=S^{-1}O\beta O^*S^{-1}.\]
Since $R_1$ is a quantum channel it follows that $Q'(S^2)=T^{-2}$.  Choose
\[\phi_T=T\alpha T, \phi_0=S^{-1}O\beta O^* S^{-1}, \hat\phi_0=S^2,\hat \phi_T=T^{-2}, \chi_0=\beta^{\frac{1}{2}}O^*S^{-1},\chi_T=\alpha^{\frac{1}{2}}T.\]
Then
\begin{eqnarray*}
&&Q(\phi_T)=\phi_0, \quad Q'(\hat \phi_0)=\hat \phi_T,\\ &&\chi_0^*\chi_0=S^{-1}O\beta^{\frac{1}{2}}\beta^{\frac{1}{2}}O^*S^{-1}=\phi_0,\quad
\chi_T^*\chi_T=T\alpha^{\frac{1}{2}}\alpha^{\frac{1}{2}}T=\phi_T,
\\
&&\rho_0=\beta=\beta^{\frac{1}{2}}O^*S^{-1} S^2 S^{-1} O \beta^{\frac{1}{2}}=\chi_0\hat \phi_0\chi_0^*,\quad \rho_T=\alpha=\alpha^{\frac{1}{2}}T T^{-2} T \alpha^{\frac{1}{2}}=
\chi_T\hat \phi_T \chi_T^*.
\end{eqnarray*}
\qed

We now note that Conjecture \ref{GPcon}, under the assumption that $\rho_T$ and $\rho_0$ are positive definite,  implies that $Q$ can be scaled to a quantum channel $R$ satisfying $R(\rho_T)=\rho_0$.  Observe first that \eqref{GPcon2} yields that $\chi_0, \chi_T$ are invertible.  Let
\[T=\chi_T^*\rho_T^{-\frac{1}{2}}, \; S=\rho_0^{\frac{1}{2}}(\chi_0^*)^{-1}, \quad  R(X)=SQ(TXT^*)S^*.\]
Then
\begin{eqnarray*}
&&R(\rho_T)=SQ(T\rho_T T^*)S^*=SQ(\chi_T^*\rho_T^{-\frac{1}{2}}\rho_T\rho_T^{-\frac{1}{2}}\chi_T)S^*=SQ(\chi_T^*\chi_T)S^*=\\
&&SQ(\phi_T)S^*=S\phi_0S^*=S\chi_0^*\chi_0 S^*=\rho_0^{\frac{1}{2}}(\chi_0^*)^{-1}\chi_0^*\chi_0 \chi_0^{-1}\rho_0^{\frac{1}{2}}=\rho_0,\\
&&R'(I_n)=T^*Q'(S^*S)T=T^*Q'(\chi_0^{-1} \rho_0(\chi_0^*)^{-1})=
T^*Q'(\hat\phi_0)T=\\
&&T^* \hat \phi_T T^*=\rho_T^{-\frac{1}{2}}\chi_T \hat \phi_T\chi_T^*\phi_T^{-\frac{1}{2}}=\rho_T^{-\frac{1}{2}}\rho_T \rho_T^{-\frac{1}{2}}=I_n.
\end{eqnarray*}
\section{Uniqueness of  fixed points}\label{sec:uniq}
Denote by $\P\rH_{n,++}$ the space of all rays $t X, X\in\rH_{n,++}, t>0$.  Clearly, we can identify $\P\rH_{n,++}$ with $\rH_{n,++,1}$.
The Hilbert metric on $\P\rH_{n,++}$ is given as follows:
\begin{equation}\label{defHilmet}
\dist(X,Y)=\log \lambda_1(XY^{-1})-\log\lambda_n(XY^{-1}), \quad X,Y\in\rH_{n,++}.
\end{equation}

To justify that $\dist(X,Y)$ is a metric on $\P\rH_{n,++}$ we recall the following facts.
\begin{equation}\label{XYinid}
\lambda_{1}(XY^{-1})=\lambda_1(Y^{-\frac{1}{2}}X Y^{-\frac{1}{2}})=\|Y^{-\frac{1}{2}}X Y^{-\frac{1}{2}}\|, 
\;\lambda_n(XY^{-1})=\frac{1}{\|Y^{\frac{1}{2}}X ^{-1}Y^{\frac{1}{2}}\|}.
\end{equation}
Hence
\begin{eqnarray*}
&&\lambda_1(XZ^{-1})=\lambda_1(Y^{-\frac{1}{2}}XY^{-\frac{1}{2}} Y^{\frac{1}{2}} Z^{-1}Y^{\frac{1}{2}})\le 
\|Y^{-\frac{1}{2}}XY^{-\frac{1}{2}} Y^{\frac{1}{2}} Z^{-1}Y^{\frac{1}{2}}\|\le\\
&&\|Y^{-\frac{1}{2}}XY^{-\frac{1}{2}}\| \|Y^{\frac{1}{2}} Z^{-1}Y^{\frac{1}{2}}\|=\lambda_1(XY^{-1})\lambda_1(YZ^{-1}).
\end{eqnarray*}
By replacing $X$, $Y$ and $Z$ by their inverses and using \eqref{Xnormid} we deduce
\[\lambda_{n}(XZ^{-1})\ge \lambda_n(XY^{-1})\lambda_n(YZ^{-1}).\]
Combine this two inequalities to deduce that  
\[0\le \dist(X,Z)\le \dist(X,Y)+\dist(X,Z) \textrm{ for } X,Y,Z\in \rH_{n,++}.\]
Suppose that $\lambda_1(XY^{-1})=\lambda_n(XY^{-1})$.  Hence $\lambda_1(Y^{-\frac{1}{2}}X Y^{-\frac{1}{2}})=\lambda_n(Y^{-\frac{1}{2}}X Y^{-\frac{1}{2}})$.
Therefore $Y^{-\frac{1}{2}}X Y^{-\frac{1}{2}}=tI\Rightarrow X=tY$ for some $t>0$. 

Assume that $\alpha\in\rH_{n,++}$.  As the maps $Q,Q', \tilde Q, D_{\alpha}$ are homogeneous maps of degrees $1,1,0,0$ it follows that the maps
$Q,Q',\tilde Q, D_{\alpha}:\rH_{n,++}\to \rH_{n,++}$ induce the corresponding maps from $\P\rH_{n,++}$ to itself.  By abusing notation we denote these
maps as $Q,Q',\tilde Q, D_{\alpha}:\P\rH_{n,++}\to \P\rH_{n,++}$ and no ambiguity will arise.  Clearly, the maps $Q'$ and $\tilde Q$ from $\P\rH_{n,++}$ to itself
are identical.  Furthermore, we identify the maps $Q,\tilde Q, D_{\alpha}:\P\rH_{n,++}\to \P\rH_{n,++}$ with the corresponding maps  $Q,\tilde Q, D_{\alpha}:\rH_{n,++,1}\to \P\rH_{n,++,1}$.  We also will view $\rH_{n,++,1}$ as metric space with respect the the Hilbert metric.
\begin{theo}\label{strictonctrPhi}  Let $Q$ be a positive quantum channel given by \eqref{defcp}.  Then
\begin{enumerate}
\item The maps $Q$ and $\tilde Q$ are strict contractions on $\P\rH_{n,++}$:
\begin{eqnarray}\label{contest1}
\dist (Q(X),Q(Y)\le \kappa \dist(X,Y), \;\dist(\tilde Q(X),\tilde Q(Y))\le \kappa \dist(X,Y),\\ 
\kappa=\frac{b(Q)-a(Q)}{b(Q)+a(Q)}.\label{contest2}
\end{eqnarray}
\item  The map $D_{t I_n}: \rH_{n,++}\to \rH_{n,++}$ preserves the Hilbert metric on $\P\rH_{n,++}$ for $t>0$.
\item The map $\Theta=\Phi_{\frac{1}{n}I_n,\frac{1}{n}I_n}:\rH_{n,++,1}\to \rH_{n,++,1}$ is a contraction with respect the Hilbert metric.  The contraction constant is bounded by $\kappa^2$.
\item $\Theta$ has a unique fixed point $F\in\rH_{n,+,1}$ which lies in the interior of $\rH_{n,+,1}$.   For each $X\in \rH_{n,+,1}$ the iterations $\Theta^{\circ m}(X), m\in\N$ converge $F$.
\item There exist open balls $B_1,B_2\subset \rH_{n,++,1}$, in the Frobenius norm, centered at $\frac{1}{n}I_n, F$, respectively, with positive radii $\varepsilon_1,\varepsilon_2$ respectively, with the following properties.  Assume that $\alpha,\beta\in B_1$. Then $\Phi_{\alpha,\beta}$ has a unique fixed point $U=U(\alpha,\beta)$ which lies in $B_2$.  Furthermore,
for each $X\in B_2$ the iterations  $\Theta^{\circ m}(X), m\in\N$ converge $U(\alpha,\beta)$.

\end{enumerate}
\end{theo}
\proof \emph{1}.  We first show the first inequality of \eqref{contest1}.  Recall that the map $Q:\rH_{n,+}\to \rH_{n,+}$ is linear.
Hence we can apply Birkhoff's theorem \cite{Bir57}, which gives an upper bound on the strict contraction of $Q$ on $\P\rH_{n,++}$.   Namely, let
\[\Delta(Q):=\max\{\dist(Q(X),Q(Y)), X,Y\in \rH_{n,+,1}\}.\]
Then 
\[\dist (Q(X),Q(Y))\le \tanh (\frac{1}{4}\Delta(Q)) \dist (X,Y) \textrm{ for all }X,Y\in \rH_{n,++}.\]
As $a(Q)I_n \preceq Q(X),Q(Y) \preceq Q(b)I_n$ it follows that $ \frac{a(Q)}{b(Q)}Q(Y)\preceq Q(X) \preceq \frac{b(Q)}{a(Q)}Q(Y) $.
Hence $\dist(Q(X),Q(Y))\le 2\log\frac{b(Q)}{a(Q)}$.  Therefore
$\Delta(Q)\le  2\log\frac{b(Q)}{a(Q)}$.  Note that
\[\tanh (\frac{1}{4}\Delta(Q))\le \tanh(\frac{1}{2}\log\frac{b(Q)}{a(Q)}) =\frac{\sqrt{\frac{b(Q)}{a(Q)}}-\sqrt{\frac{a(Q)}{b(Q)}}}{\sqrt{\frac{b(Q)}{a(Q)}}+\sqrt{\frac{a(Q)}{b(Q)}}}=\frac{\frac{b(Q)}{a(Q)}-1}{\frac{b(Q)}{a(Q)}+1}=\frac{b(Q)-a(Q)}{b(Q)+a(Q)}.\]
This shows the first part of the inequality \eqref{contest1}.

Consider $\tilde Q:\P\rH_{n,++}\to \P\rH_{n,++}$.  Clearly, $\dist(\tilde Q(X),\tilde Q(Y))=\dist(Q'(X),Q'(Y))$.  Hence, it is enough to estimate the contraction of $Q'$.
Apply Lemma \ref{posQprime}  and the above arguments to deduce the second part of the inequality \eqref{contest1}.

\noindent
\emph{2}.
Let $\Omega:\rH_{n,++}\to \rH_{n,++}$ be given by $\Omega(X)=X^{-1}$.
Clearly, $\dist(D_{t I_n}(X),D_{t I_n}(Y))=\dist(\Omega(X),\Omega(Y))$ for $X,Y\in \rH_{n,++}$.  Observe that
\[\lambda_1(\Omega(X)\Omega(Y)^{-1})=\lambda_n(XY^{-1})^{-1}, \; \lambda_n(\Omega(X)\Omega(Y)^{-1})=\lambda_1(XY^{-1})^{-1}\]
for $X,Y\in \rH_{n,++}$.
Hence $D_{tI_n}$ preserves the metric on $\rH_{n,++}$ for $t>0$.

\noindent
\emph{3} follows straightforward from \emph{2} and \emph{3}.

\noindent
\emph{4}.  As $\Theta(\rH_{n,+,1})\subset \rH_{n,++,1}$ and $\Theta$ is a contraction on $\rH_{n,++,1}$, Banach Fixed Point Theorem yields that $F\in \rH_{n,++,1}$
is a unique fixed point of $\Theta$.  Furthermore, for each $X\in \rH_{n,+,1}$ the iterations $\Theta^{\circ m}(X), m\in\N$ converge $F$.

\noindent
\emph{5}.  Denote by $\cS_n,\cA_n\subset \R^{n\times n}$ the space of $n\times n$ real symmetric and real skew symmetric matrices.
Note that $\dim\cS_n=\frac{n(n+1)}{2}$ and $\dim \cA_n=\frac{n(n-1)}{2}$.
We now consider $X\in \rH_{n}$ as of the form $X=X_{\Re}+\i X_{\Im}$, where $X_{\Re}\in\cS_n,X_{\Im}\in\cA_n$.
That is we view $\rH_{n}$ as the space $\R^{n^2}$, where the $n^2$ variables are the real entries of the upper diagonal and strict upper diagonal parts of $X_{\Re}$
and $X_{\Im}$ respectively.
An open ball of radius $r> 0$ centered at $X_0\in \rH_n$ is $B(X_0,r)=\{X_0+W,\;W\in \rH_{n}, \tr W^2< r^2 \}$.
 Thus $\rH_{n,++}$ is a domain in $\rH_{n}$.
A map $\Lambda: \rH_{n,++}\to \rH_{n,++}$ is called real analytic if the following conditions holds.   Let  $X_0\in\rH_{n,++}$ be given.
Then there exists an open ball centered at $X_0$ of radius $\varepsilon=\varepsilon(X_0,\Lambda)$, which is contained in $\rH_{n,++}$, such that the following conditions hold.  For
 $X=X_0+W$ in this ball the real and the imaginary parts of the entries of $\Lambda(X_0+W)$ are given by convergent power series in the real and complex parts of the entries of $W$.  Similar definition applies for analyticity of the map  $\Lambda_1:\rH_{n,++,1}\to \rH_{n,++,1}$. 

Clearly, the maps $Q,Q':\rH_{n,++}\to \rH_{n,++}$ are linear in $n^2$ real variables of $X\in \rH_{n,++}$.  Hence these maps are analytic.  It is straightforward to see that the map
$\tilde Q:\rH_{n,++}\to \rH_{n,++,1}\subset \rH_{n,++}$ is analytic.  Use the formula for $\Omega(X)$ in terms of the adjoint matrix of $X$, i.e., $\Omega(X)=\frac{1}{\det X} \adj X$, to deduce that $\Omega$ is analytic on
$\rH_{n,++}$.  

Let $\Gamma:\rH_{n,++}\to \rH_{n,++}$ be given by $\Gamma(X)=X^{\frac{1}{2}}$.  Then $\Gamma$ is analytic on $\rH_{n,++}$.
This fact can be easily deduced from the Cauchy integral formula \cite[\S3.4]{Frbk}.  Indeed, let $\C_+=\{z\in \C, \Im z>0\}$ be the right half complex plane.  Let $f:\C_+\to \C_+$ be the 
analytic function $f(z)=\sqrt{z}$, ($\sqrt{1}=1$).  Assume that $X_0\in \rH_{n,++}$ is given.   Let $S$ be the circle centered at $\frac{\lambda_1(X_0)+\lambda_n(X_0)}{2}$
with radius $\frac{\lambda_1(X_0)}{2}$.  Assume that $\varepsilon>0$ is small enough so that $\lambda_n(X_0+W)>\frac{\lambda_n(X_0)}{2}$ for each $W\in B(0,\varepsilon)$.
Then
\[\Gamma(X_0+W)=\frac{1}{2\pi\i}\int _S f(z)(zI-(X_0+W))^{-1}dz.\]
 Express $(zI-(X_0+W))^{-1}$	in terms of the adjoint of $(zI-(X_0+W))$ to deduce the analyticity of $\Gamma$.

Hence the map  $D_{\alpha}:\rH_{n,++}\to \rH_{n,++,1}$ is analytic for $\alpha\in \rH_{n,++}$.  Let $\Phi_{\alpha,\beta}$ be the map given by \eqref{defPhi}.
Then $\Phi_{\alpha,\beta}:\rH_{n,++,1}\to \rH_{n,++,1}$ is analytic.  In particular,  $\Theta$ is analytic on $\rH_{n,++,1}$.  We now consider $\Theta$ in the neighborhood
of the fixed point $F\in\rH_{n,++,1}$.  Denote by $\rH_{n,0}=\{W\in\rH_n, \tr W=0\}$.  Then the Jacobian of $\Theta|\rH_{n,++,1}$ at $F$ is a linear map $P:\rH_{n,0}\to \rH_{n,0}$.  
That is 
\[\Theta(F+W)=F+P(W)+ \textrm{ higher order terms}.\]
Denote by $\rho(P)$ the spectral radius $P$.  We claim that $\rho(P)<1$.    By considering a scaled quantum channel, we may assume without loss of generality that $Q$ is a unital channel, i.e., $Q(\frac{1}{n}I_n)=\frac{1}{n}I_n$.   Hence $F=\frac{1}{n}I_n$.
Assume that $W\in  \rH_{n,0}$ and consider $F(t)= \frac{1}{n}I+tW$.  Then
\begin{eqnarray*}
&&Q(F(t))=\frac{1}{n}I+tQ(W),\;D_{\frac{1}{n}I_n}(\frac{1}{n}I+tQ(W))=\frac{1}{n}I_n-tQ(W)+O(t^2),\\
&&\tilde Q(\frac{1}{n}-tQ(W)+O(t^2))=\frac{1}{n}I_n-t(Q'\circ Q)(W))+O(t^2),\\
&&D_{\frac{1}{n}I_n}(\frac{1}{n}I-t(Q'\circ Q)(W)+O(t^2))=\frac{1}{n}I_n+t(Q'\circ Q)(W)+O(t^2).
\end{eqnarray*}

Therefore $P=Q'\circ Q$ is a selfadjoint linear operator, which is positive semi-definite with respect to the standard inner product on $\rH_{n,0}$.
That is, all eigenvalues of $P$ are nonnegative. We claim that $Q$ and $Q'$ are strict contraction on $\rH_{n,0}$.  Indeed $Q(\rH_{n,+,1})\subset \rH_n(a(Q),b(Q),1)$
and $Q(\frac{1}{n}I_n)=\frac{1}{n}I_n$.  Hence any ball in $\rH_{n,+,1}$ centered at $\frac{1}{n}I_n$ is mapped to the interior of this ball.  Similarly, for $Q'$.
Hence $P$ is a positive semi-define matrix with $\|P\|=\rho(P)<1$.  In particular $P$ is a contraction on $\rH_{n,0}$ with respect to the Frobenius norm.

Consider now $\Phi_{\alpha,\beta}$ as the family of maps depending on parameters $\alpha,\beta$.  That is, $\Phi_{\alpha,\beta}$ is given by $(X,\alpha,\beta)\mapsto \Phi_{\alpha,\beta}(X)$ for 
$(X,\alpha,\beta)\in \rH_{n,++}^3$.  The arguments above show that $\Phi_{\alpha,\beta}$ is analytic on  $\rH_{n,++}^3$.  Hence, in the neighborhood of
$(\frac{1}{n}I_n,\frac{1}{n}I_n,\frac{1}{n}I_n)$ in $\rH_{n,++,1}^3$ we have:
\begin{eqnarray*}
&&\Phi_{\frac{1}{n}I_n+\alpha_1,\frac{1}{n}I_n+\beta_1}(\frac{1}{n}I_n+W)=\\
&&\frac{1}{n}I_n+W_1(\alpha_1,\beta_1)+
(P(W)+E_1(\alpha_1,\beta_1)(W))+E_2(W,\alpha_1,\beta_1),\\
&&\|W_1(\alpha_1,\beta_1)\|_F\le K(\|\alpha_1\|_F+\|\beta_1\|_F), \|E_1(\alpha_1,\beta_1)\|\le K(\|\alpha_1\|_F+\|\beta_1\|_F),\\
&&E_2(0,\alpha_1,\beta_1)=0,\;\|E_2(W_1,\alpha_1,\beta_1)-E_2(W_2,\alpha_1,\beta_1)\|_F\le \\
&&K\|W_1-W_2\|_F(\|W_1\|_F+\|W_2\|_F) (1+\|\alpha_1\|_F+\|\beta_1\|_F).
\end{eqnarray*}
Note that $W_1(\alpha_1,\beta_1), E_2(W,\alpha_1,\beta_1)\in \rH_{n,0}$ and $E_1(\alpha_1,\beta_1):\rH_{n,0}\to \rH_{n,0}$ is a linear operator for each $\alpha_1,\beta_1$
in some open ball of radius $r$ centered at $0\in\rH_{n,0}$.  Furthermore, we assume that the above inequalities hold for $\|\alpha_1\|_F,\|\beta_1\|_F,\|W\|,\|W_1\|,\|W_2\|<r$.
Assume that $\|P\|<1-2\eta$, for some $\eta>0$, and $t\in (0,r)$.  Suppose that $\|\alpha_1\|,\|\beta_1\|<\frac{t^2}{2}$ and $\|W\|_F<t$.
Then
\[\|\Phi_{\frac{1}{n}I_n+\alpha_1,\frac{1}{n}I_n+\beta_1}(\frac{1}{n}I_n+W)-\frac{1}{n}I_n\|< Kt^2+(1-2\eta +Kt^2)t+Kt^2(1+t^2). \]
Hence there exists $\varepsilon \in (0,\min(r,1))$ such that 
\[2K(\varepsilon+\varepsilon^2 +\varepsilon^3)<\eta.\]
In particular
\[K\varepsilon^2+(1-2\eta+K\varepsilon^2)\varepsilon+K\varepsilon^2(1+\varepsilon^2)=(1-2\eta+K(2\varepsilon+\varepsilon^2+\varepsilon^3))\varepsilon<(1-\eta)\varepsilon.\]
Therefore
\[\Phi_{\alpha,\beta}(B(\frac{1}{n}I_n,\varepsilon))\subset B(\frac{1}{n}I_n,\varepsilon) \textrm{ for } \alpha,\beta\in B(\frac{1}{n}I_n,\frac{\varepsilon^2}{2}).\]
Assume that $W_1,W_2\in B(0,\varepsilon)$.  Then
\begin{eqnarray*}
&&\|\Phi_{\alpha,\beta}(\frac{1}{n}I_n+W_1)-\Phi_{\alpha,\beta}(\frac{1}{n}I_n+W_2)\|_F\le \\
&&(1-2\eta+K\varepsilon^2)\|W_1-W_2\|+2K\varepsilon(1+\varepsilon^2)\|W_1-W_2\|=\\
&&(1-2\eta+K(2\varepsilon+\varepsilon^2 +2\varepsilon^3))\|W_1-W_2<(1-\eta)\|W_1-W_2\|.
\end{eqnarray*}
That is, for fixed $\alpha,\beta \in  B(\frac{1}{n}I_n,\frac{\varepsilon^2}{2})$ and $X\in B_2=B(\frac{1}{n}I_n,\varepsilon)$ the map $\Phi_{\alpha,\beta}$ is a contraction with respect to
the Frobenius norm.  Hence $\Phi_{\alpha,\beta}$ has a unique fixed point $U=U(\alpha,\beta)$ in $ B_2$.  Furthermore,
for each $X\in  B_2$ the iterations  $\Theta^{\circ m}(X), m\in\N$ converge to $U$.

It is left to show that there exists $\varepsilon_1\in (0,\frac{\varepsilon^2}{2})$ such that for $\alpha,\beta \in  B(\frac{1}{n}I_n,\varepsilon_1)$ the map $\Phi_{\alpha,\beta}$
have a unique fixed point which is $U(\alpha,\beta)$.  Since $\Theta$ has a unique fixed point $\frac{1}{n}I_n$ the continuity argument yields that there exists  
$\varepsilon_1\in (0,\frac{\varepsilon^2}{2})$ such that for each $\alpha,\beta\in  B_1=B(\frac{1}{n}I_n,\varepsilon_1)$ all fixed points of  $\Phi_{\alpha,\beta}$ lie in  $B_2$.   Our previous results show that in $B_2$ the map $\Phi_{\alpha,\beta}$ has a unique fixed point $U(\alpha,\beta)$.\qed

It is an open problem to prove or give a counterexample that $\Phi_{\alpha,\beta}$ has a unique fixed point in $\rH_{n,++,1}$ for each $\alpha,\beta\in \rH_{n,++,1}$.
See \cite{GP15} for numerical simulations.

\bibliographystyle{plain}

\end{document}